\documentclass[a4paper,11pt]{article}
\usepackage{pos}
\usepackage{braket}
\usepackage{dsfont}
\usepackage{subcaption}

\newcommand{\de}{{\rm d}}
\newcommand{\be}{\mathbf{b}}

\newcommand{\bvec}[1]{\mathbf{#1}}

\newcommand{\vecb}{\bvec b}

\newcommand{\absb}{|\vecb|}

\newcommand{\lrb}[1]{\left( #1 \right)}
\newcommand{\lrsb}[1]{\left[ #1 \right]}
\newcommand{\lrcb}[1]{\left\{ #1 \right\}}

\renewcommand{\Re}[1]{\operatorname{Re}\lrsb{#1}}
\renewcommand{\Im}[1]{\operatorname{Im}\lrsb{#1}}

\title{Critical Unstable Qubits in Particle Physics}

\author[a,b]{D. Karamitros}
\author[a]{T. McKelvey}
\author[a]{S. Panghal}
\author*[a,c]{A. Pilaftsis}

\affiliation[a]{Department of Physics and Astronomy, University of Manchester,\newline
Manchester, M13 9PL, United Kingdom
}

\affiliation[b]{Dipartimento di Fisica e Astronomia, Universit\`{a} degli Studi di Padova,\newline
Via Marzolo 8, 35131 Padova, Italy}

\affiliation[c]{PRISMA Cluster of Excellence \& Mainz Institute for Theoretical Physics,\newline
Johannes Gutenberg University, 55099 Mainz, Germany}

\emailAdd{dimitros.karamitros@paduaemail.com}
\emailAdd{thomas\_mckelvey@outlook.com}
\emailAdd{snehit.panghal@postgrad.manchester.ac.uk}
\emailAdd{apostolos.pilaftsis@manchester.ac.uk}

\abstract{We study in detail the dynamics of unstable two-level quantum systems by adopting the Bloch-vector representation. We identify a novel class of critical scenarios in which the so-called energy-level and decay-width vectors, ${\bf E}$ and ${\bf\Gamma}$, are orthogonal to one another, and the parameter $r = |{\bf \Gamma}|/(2|{\bf E}|)$ is less than~1. Most remarkably, we find that critical unstable qubit systems exhibit atypical behaviours like coherence--decoherence oscillations when analysed in an appropriately defined co-decaying frame of the system. By making use of a Fourier series decomposition, we define anharmonicity observables that quantify the degree of non-sinusoidal oscillation of a CUQ. We apply the results of our formalism to the neutral-meson systems and derive generic upper limits on these new observables. In particular, we provide a compilation table of all well-explored meson--antimeson two-level systems in terms of Bloch-sphere parameters.}

\FullConference{Corfu Summer Institute 2025 ``Workshop on the Standard Model and Beyond", August
24 - September 3, 2025, Corfu, Greece.}

\begin{document}
\maketitle

\section{Introduction}

Quantum entanglement and quantum coherence are two topics in Quantum Mechanics~(QM) that received much attention in several modern physics explorations. They have extensively been tested in terrestrial experiments, and suggestions have been made recently to probe them in satellite missions in space~\cite{Alonso:2022oot}. These two concepts of~QM play also a central r\^{o}le in our understanding of quantum systems, including elementary systems such as two-level quantum systems, also known as {\it qubits}~\cite{PhysRevA.51.2738}.

Historically, two-level quantum systems were introduced in 1937 by Rabi~\cite{Rabi:1937dgo} to  describe his nominal oscillatory phenomena such as spin precession of fermions in the presence of magnetic fields, with  notable applications to nuclear magnetic resonance and medicine. In the context of particle physics, these ideas have been reinvented by Lee, Oehme and Yang (LOY)~\cite{Lee:1957qq} to address possible phenomena of charge (C) and charge-parity (CP) violation in the kaon system through $K^0\bar{K}^0$ oscillations.

If the two-level quantum system turns out to be unstable, like the kaon system mentioned above, one has to then consider decay-width effects in addition to oscillations. These effects have been taken   into account by LOY~\cite{Lee:1957qq}, who made use of the famous
effective approximation developed in the 1930s by Weisskopf and Wigner~(WW)~\cite{Weisskopf1930}. In the WW approximation, the dynamics of an unstable quantum system may be conveniently described by using an effective Hamiltonian, $\textrm{H}_\textrm{eff}$.
The time evolution of an unstable quantum state, $\ket{\Psi}$, is then governed by an effective Schr\"{o}dinger equation,
\begin{equation}
    i\partial_t \ket{\Psi} = \textrm{H}_\textrm{eff} \ket{\Psi}.
    \label{eq:Schrodinger_def}
\end{equation}
At sufficiently large times, an initially mixed unstable qubit is expected to approach a pure QM state aligned to its long-lived eigenstate of ${\rm H}_{\rm eff}$. However, if the two eigenstates of ${\rm H}_{\rm eff}$ happen to have equal lifetimes, the evolution of quantum coherence for the unstable qubit will become rather complex. In order to accurately capture all the features emanating from such dynamics, we will adopt in our study the Bloch-sphere formalism of qubits~\cite{PhysRev.70.460}. In particular, we will employ the Bloch vector representation to decompose the non-Hermitian Hamiltonian ${\rm H}_{\rm eff}$ in terms of the energy-level and decay-width four-vectors, $E^\mu = (E^0, {\bf E})$ and $\Gamma^\mu = (\Gamma^0,{\bf\Gamma})$, respectively.

\section{The Density Matrix}\label{sec:Bloch}

The density matrix formalism of quantum mechanics~\cite{PhysRev.70.460, Sakurai:1341875} offers a simple extension it the standard paradigm by permitting the study of mixed state systems on top of the well-studied pure state systems. The density matrix operator, $\rho$, is defined using the complete collection of normalised eigenstates of the Hamiltonian, $\ket{\Psi^k}$, as the weighted sum of projection operators:
\begin{equation}\label{eq:denmat}
    \rho \: = \: \sum_{k} \, w_k \, \ket{\Psi^k}\bra{\Psi^k} \, .
\end{equation}
Where the weights $w_k$ are positively valued and normalised such that the sum over all $w_k$ is equal to one. Finding the expectation of the density matrix along a particular eigenstate of the Hamiltonian $\ket{\Psi^k}$, one can define $w_k$ to be equivalent to the probability of the state being observed in this eigenstate, i.e. $\mathbb{P}\left(\Psi^k \right)$. Further to this, when acting on a generic quantum state, $\rho$ segments the state along each of the Hamiltonian eigenstates, and when traced over alongside a generic operator, gives the expectation value of that operator. This interpretation of the density matrix as a statistical object as well as its manifest structure provides a number of conditions which must be satisfied. First, the expectation value of the identity operator must return unity, demonstrating the completeness of the space:
\begin{equation}
    {\rm Tr} \, \rho \: = \: 1 \, .
\end{equation}
Second, since $w_k$ are to be interpreted as probabilities, these quantities are real valued, and so $\rho$ is Hermitian, $\rho^\dagger = \rho$. Finally, we can see that in the case that all the probability is isolated into a single eigenstate, i.e. $w_1 = 1$ and $w_{k>1} = 0$, the density matrix takes the concise form:
\begin{equation}
    \rho \: = \: \ket{\Psi^1}\!\bra{\Psi^1} \, ,
\end{equation}
and satisfies the idempotency condition $\rho^2 = \rho$. It can be demonstrated that this condition cannot hold for general constructions of the density matrix, and so we can simply define the purity of a state using this idempotency condition. Wherever $\rho^2 = \rho$, the density matrix describes a pure state, otherwise, the state is mixed. 

The density matrix is a well-established description of the dynamics of unitary systems, where the time-dependency of the states is determined through a Schr\"{o}dinger-like evolution using a Hermitian Hamiltonian, for example the polarisation of light~\cite{PoincareH1889}, or spin states in magnetic fields~\cite{PhysRev.70.460}. In contrast, the literature is less complete in the study of unstable or open systems. The evolution of these systems makes use of a so-called effective Hamiltonian:
\begin{equation}
    {\rm H}_{\rm eff} \: = \: {\rm E} - \frac{i}{2}\Gamma \, ,
\end{equation}
where ${\rm E}$ and $\Gamma$ are Hermitian matrices that describe dispersive and dissipative effects, respectively. Making use of this effective Hamiltonian, we can show that the density matrix evolves according to:
\begin{equation}
    \frac{{\rm d} \rho}{{\rm d} t} \: = \: -i {\rm H}_{\rm eff}\rho +i \rho {\rm H}_{\rm eff}^\dagger \: = \:  -i [{\rm E} \, , \, \rho] \, - \, \frac{1}{2} \left\{ \Gamma \, , \, \rho \right\} \, .
\end{equation}
Observe that in this extended case, the Hermiticity of the density matrix is preserved; however, the unit trace is lost as the system evolves. Inspection of the statistical structure finds an inconsistency since probability is flowing out of the density matrix:
\begin{equation}
    \frac{{\rm d} {\rm Tr} \, \rho}{{\rm d} t} \: = \: -\,{\rm Tr} \left( \rho \Gamma \right) \: \neq \: 0 \, .
\end{equation}
This can be easily resolved by noticing that the density matrix we have been considering thus far is simply a sub-matrix of a wider density matrix, and that the Hilbert space we are dealing with is a sub-space of a larger Hilbert space containing not only the parent space $\mathcal{H}_P$, but also a daughter space, $\mathcal{H}_D$, i.e. $\mathcal{H} = \mathcal{H}_P \otimes \mathcal{H}_D$. It is, therefore, necessary to adjust the density matrix such that we only consider the parent Hilbert space and restore a unitary description. This is achieved simply by normalising relative to the trace:
\begin{equation}\label{eq:NormDenMat}
    \widehat{\rho} \: = \: \frac{\rho}{{\rm Tr} \, \rho} \, .
\end{equation}
Making use of the evolution equations already presented, one can verify that the normalised density matrix evolves according to the expression:
\begin{equation}\label{eq:NormDenMatEvol}
    \frac{{\rm d} \widehat{\rho}}{{\rm d} t} \: = \:  -i [{\rm E} \, , \, \widehat{\rho}] \, - \, \frac{1}{2} \left\{ \Gamma \, , \, \widehat{\rho} \right\} \, + \, \widehat{\rho} \, {\rm Tr} \left( \widehat{\rho} \Gamma \right) \, .
\end{equation}
This modified evolution equation is manifestly probability preserving, as evidenced by it being traceless provided that one makes the natural assumption ${\rm Tr} \, \rho = 1$. However, this correction comes at the cost of linearity. In the case of non-trivial decay processes, i.e. $\Gamma \not\propto \mathds{1}_2$, it is clear that the evolution equation is quadratic in $\widehat{\rho}$. We will see that these non-trivialities give rise to novel phenomena which thus far have been left unexplored.

For the two-level system, otherwise referred to as a \textit{qubit} system, we consider a parametrization of the density matrix which further simplifies the dynamics into a geometric description. The Bloch sphere representation of the density matrix offers a concise way of characterising the full information of the state, including its coherences, using a single vector, $\mathbf{b} \in \mathbb{R}^3$, the so-called Bloch vector. Since the density matrix is Hermitian, it may be expressed on the Pauli basis, with the zeroth component determined to be one to ensure the density matrix is unitary:
\begin{equation}
    \widehat{\rho} \: = \: \frac{1}{2} \left(\mathds{1}_2 \, + \, \mathbf{b}\cdot\boldsymbol{\sigma} \right) \, .
\end{equation}
Henceforth, $\mathbf{b}$ shall be referred to as the co-decaying Bloch vector to mirror its origin in the normalised density matrix. Making use of the coherence condition $\rho^2 = \rho$, one can verify that a pure state may be characterised as one where $|\mathbf{b}| = 1$, and a mixed state $|\mathbf{b}|<1$.

Similarly, both the energy level matrix ${\rm E}$ and decay matrix $\Gamma$ can be expressed over the Pauli basis:
\begin{equation}
    {\rm E} \: = \: {\rm E}^0 \mathds{1}_2 \, - \, \mathbf{E}\cdot \boldsymbol{\sigma} \, , \qquad \Gamma \: = \: \Gamma^0 \mathds{1}_2 \, - \, \mathbf{\Gamma}\cdot \boldsymbol{\sigma} \, .
\end{equation}
Inputting these representations into equation~\eqref{eq:NormDenMatEvol}, we derive the evolution equation of the co-decaying Bloch vector by taking the coefficients of each Pauli basis matrix:
\begin{equation}\label{eq:bEvolEq1}
    \frac{{\rm d} \mathbf{b}}{{\rm d}t} \: = \: -2 \mathbf{E} \times \mathbf{b} + \mathbf{\Gamma} - \left( \mathbf{b} \cdot \mathbf{\Gamma} \right) \mathbf{b} \, .
\end{equation}
We see that the trivial components, ${\rm E}^0$ and $\Gamma^0$, drop out due to the normalisation, leaving the dynamics of $\mathbf{b}$ to be entirely determined through the energy and decay vectors, $\mathbf{E}$ and $\mathbf{\Gamma}$, respectively. For the sake of simplicity, it is convenient to introduce two dimensionless parameters:
\begin{equation}
    \tau \: = \: |\mathbf{\Gamma}|t \, , \qquad r \: = \: \frac{|\mathbf{\Gamma}|}{2|\mathbf{E}|} \, ,
\end{equation}
as well as the unit vectors $\mathbf{e} = \mathbf{E}/|\mathbf{E}|$ and $\boldsymbol{\gamma} = \mathbf{\Gamma}/|\mathbf{\Gamma}|$. Using these, equation~\eqref{eq:bEvolEq1} may be recast into the dimensionless form:
\begin{equation}\label{eq:bEvolEq2}
    \frac{{\rm d} \mathbf{b}}{{\rm d}\tau} \: = \:  - \frac{1}{r} \mathbf{e} \times \mathbf{b} + \boldsymbol{\gamma} - \left( \mathbf{b} \cdot \boldsymbol{\gamma} \right) \mathbf{b} \, .
\end{equation}
This differential equation, dependent only on two unit vectors and the dimensionless parameter $r$, is the \textit{master evolution equation} of $\mathbf{b}$, and forms the basis of our analysis on the behaviour of unstable qubit systems.

In principle, the decay processes should drive the co-decaying Bloch vector, $\mathbf{b}(\tau)$, towards a stationary solution, i.e. the longest lived state of the qubit system. By projecting the motion of $\mathbf{b}$ back along $\mathbf{b}$ itself, one sees how the magnitude of $|\mathbf{b}|$ evolves with $\tau$:
\begin{equation}
    \frac{{\rm d} |\mathbf{b}|^2}{{\rm d}\tau} \: = \: 2\left(\boldsymbol{\gamma}\cdot\mathbf{b} \right) \left[ 1 - |\mathbf{b}|^2 \right] \, .
\end{equation}
From this, we infer that should stationary states exist, then indeed these states must be pure, even if the state begins as a fully mixed state, i.e. $\mathbf{b}(0) = \mathbf{0}$. It is, therefore, pertinent to identify scenarios in which stationary solutions exist, or perhaps more interestingly, when these \textit{do not} exist.

To this end, we extract the longest lived state by considering different geometric regimes. The general asymptotic solution of equation~\eqref{eq:bEvolEq2} may be found by making use of a Gram-Schmidt algorithm to build out the basis of the vector space in terms of the model objects $\mathbf{e}$ and $\boldsymbol{\gamma}$. On this we propose the ansatz:
\begin{equation}
    \mathbf{b}(\tau) \: = \: \alpha(\tau) \, \mathbf{e} \: + \: \beta(\tau) \, \frac{1}{s_\gamma}\mathbf{e}\times\boldsymbol{\gamma} \: + \: \eta(\tau) \, \frac{1}{s_\gamma}\mathbf{e} \times \left(\mathbf{e}\times\boldsymbol{\gamma}\right) \, .
\end{equation}
Where in this expression, $s_\gamma = \sin \theta_{\mathbf{e}\boldsymbol{\gamma}}$, and $\theta_{\mathbf{e}\boldsymbol{\gamma}} = \angle (\mathbf{e}, \boldsymbol{\gamma})$ is the angle between $\mathbf{e}$ and $\boldsymbol{\gamma}$. Passing this ansatz to the evolution equation of the co-decaying Bloch vector, we find a set of flow equations for each of the undetermined coefficients. Solving these flow equations for the asymptotic state, one can find that the general asymptotic state is found to be:
\begin{equation}
    \mathbf{b}_\star \: \equiv \: \mathbf{b}(\infty) \: = \: \alpha(\tau) \, \mathbf{e} \: - \: \frac{1}{s_\gamma^2 r}\left( 1 - \alpha^2 \right) \mathbf{e}\times\boldsymbol{\gamma} \: - \: \frac{c_\gamma}{s_\gamma^2 \alpha} \left( 1 - \alpha^2 \right) \mathbf{e} \times \left(\mathbf{e}\times\boldsymbol{\gamma}\right) \, ,
\end{equation}
where $\alpha$ is determined to be
\begin{equation}
    \alpha = \frac{{\rm sign}(c_\gamma)}{\sqrt{2}} \, \sqrt{1-r^2 + \sqrt{(1-r^2)^2 + 4c_\gamma^2 r^2}}\; .
\end{equation}

There are, however, two special cases to be considered, ($i$) when $\boldsymbol{\gamma} = \pm \mathbf{e}$, and ($ii$) when $\mathbf{e}\perp\boldsymbol{\gamma}.$ In the former case, the Gram-Schmidt algorithm fails since the two model vectors are linearly dependent. This said, careful analysis of equation~\eqref{eq:bEvolEq2} shows that the stationary solution is found when $\mathbf{e}\times\mathbf{b} = 0$, and so $\mathbf{b}$ approaches $\mathbf{e}$ as $\tau \to \infty$. In the general solution, one can see that this solution is found in the limit $s_\gamma \to 0$. In principle, this solution is a trivial solution, since it corresponds to an effective Hamiltonian which can be diagonalised using unitary transformations, and hence the system is fully decoupled in the mass basis.

In the latter case, we  can build a complete vector space using the unit vectors $\mathbf{e}$ and $\boldsymbol{\gamma}$ without relying upon the Gram-Schmidt algorithm, and so we represent the solution on the simpler basis $\left(\mathbf{e}, \, \boldsymbol{\gamma}, \, \mathbf{e}\times\boldsymbol{\gamma}\right)$. On this basis, one can follow an identical process to find the solution:
\begin{equation}
    \mathbf{b}_\star \: = \: \frac{\sqrt{r^2 - 1}}{r} \, \boldsymbol{\gamma} \: - \: \frac{1}{r} \mathbf{e}\times\boldsymbol{\gamma} \, .
\end{equation}
Clearly, in this expression, it is necessary that $r \geq 1$, since otherwise, the co-decaying Bloch vector would take on complex values, in contradiction with its construction. Consequently, we observe a Hopf bifurcation of the
perpendicular system at $r = 1$. For a CPT invariant effective Hamiltonian, $H_{\rm eff}$,
the location of this bifurcation would correspond to a two-level system having a non-diagonalisable effective Hamiltonian. We consequently distinguish this unique set of solutions where $\mathbf{e}\perp\boldsymbol{\gamma}$ and $r<1$ from the general solutions, and henceforth refer to these as \textit{critical} unstable qubits.

\section{Critical Unstable Qubits}

\subsection{Exact Solution of the CUQ}

Here, we will highlight the exact solution of CUQs. As was demonstrated in~\cite{Karamitros:2025azy}, all CUQ systems are related through a rescaling of the Bloch vector. Therefore, we may solely consider the simple case where the co-decaying Bloch vector is pure and exists on the plane spanned by $\boldsymbol{\gamma}$ and $\mathbf{e}\times\boldsymbol{\gamma}$.

It was shown in~\cite{Karamitros:2022oew} that $\mathbf{b}(\tau)$ is necessarily contained in a plane due to having vanishing torsion. By choosing to study the motion of $\mathbf{b}(\tau)$ on this plane, we may reduce the vector equation to a scalar equation by projecting along $\boldsymbol{\gamma}$, without losing information. Hence, we find the evolution equation,
\begin{equation}
   \label{eq:dbgammadtau}
    \frac{{\rm d} (\mathbf{b}\cdot\boldsymbol{\gamma})}{{\rm d}\tau} \: = \: - \frac{1}{r}\lrb{\mathbf{e}\times \mathbf{b}}\cdot\boldsymbol{\gamma} + 1 - (\mathbf{b} \cdot \boldsymbol{\gamma})^2 \, .
\end{equation}
Assuming a pure state, $|\mathbf{b}|=1$, the projection along $\boldsymbol{\gamma}$ may be written as $\mathbf{b}\cdot\boldsymbol{\gamma}=\cos\, \varphi$. For~later convenience, we redefine the angle $\varphi$ by shifting its value by $\pi/2$, such that~$\varphi = \theta + \frac{\pi}{2}$. After making this shift, the projection, $\mathbf{b}\cdot\boldsymbol{\gamma}$, becomes an odd function of $\theta$, since~
\begin{equation}
  \label{eq:bgamma}
\mathbf{b}\cdot\boldsymbol{\gamma}\: =\: -\,\sin \theta\,.
\end{equation}
Also, we employ the triple product identity to obtain the projection along $\mathbf{e}\times\boldsymbol{\gamma}$,
\begin{equation}
  \label{eq:ebgamma}
    \lrb{\mathbf{e}\times \mathbf{b}}\cdot\boldsymbol{\gamma} \: = \: -\lrb{\mathbf{e}\times \boldsymbol{\gamma}}\cdot\mathbf{b} \: = \: -\cos \, \theta \, .
\end{equation}
Substituting equation~\eqref{eq:bgamma} and~\eqref{eq:ebgamma} in equation~\eqref{eq:dbgammadtau}, we find a rather simple expression for the angular velocity, ${\rm d}\theta (\tau)/{\rm d}\tau$, given by 
\begin{equation}
  \label{eq:dthetadtau}
    \frac{{\rm d} \theta}{{\rm d} \tau} \: = \: - \frac{1}{r}\, -\, \cos \, \theta \; .
\end{equation}

Since CUQ oscillations are periodic, the choice of initial condition is arbitrary, and the equivalence of CUQs allows us to choose $|\vecb|=1$. Therefore, we will take an initial condition that provides us useful symmetries and set the initial state of ${\bf b}(\tau)$ to be along $\mathbf{e}\times\boldsymbol{\gamma}$, i.e.~$\theta(\tau=0) = 0$.  
Given that the differential equation in~\eqref{eq:dthetadtau} is separable, its analytic solution can be obtained as follows:
\begin{equation}
  \label{eq:tautheta}
       \tau \: = \: - \int_0^\theta \, \frac{r}{1+r\cos \, x}\: {\rm d} x \: = \: - \frac{2r}{\sqrt{1-r^2}} \tan^{-1}\lrb{\sqrt{\frac{1-r}{1+r}} \, \tan \, \frac{\theta}{2}} \,. 
\end{equation}
In the co-decaying frame, the oscillation period $\widehat{\rm P}$ of a CUQ may be found by integrating~\eqref{eq:dthetadtau} over the full range of $\theta \in (-\pi, \pi)$,
\begin{equation}\label{eq:DLPeriod}
    \widehat{\rm P}\: =\: \int_{-\pi}^\pi \, \frac{r}{1+r\cos \, x}\: {\rm d} x \: = \: \frac{2\pi r}{\sqrt{1-r^2}}\ .
\end{equation}
Given~$\widehat{\rm P}$,  we may define the dimensionless angular frequency
\begin{equation}
    \widehat{\omega} \: = \: \frac{2\pi}{\widehat{\rm P}} \: = \: \frac{\sqrt{1-r^2}}{r} \, .
\end{equation}
Taking this last expression into account, we find an expression for the angle $\theta$ in terms of~$\tau$,
\begin{equation}\label{eq:AngSol}
    \tan \frac{\theta}{2} \: = \: -\,\sqrt{\frac{1+r}{1-r}} \, \tan \frac{\widehat{\omega} \tau}{2} \, .
\end{equation}
In the limit $r \to 0$, we can see that equation~\eqref{eq:AngSol} returns the expected result corresponding to Rabi oscillations: $\theta (\tau) = -\widehat{\omega} \tau$, where the lingering pre-factor $-1$ indicates that ${\bf b}(\tau)$ rotates in the plane in a clockwise direction. Restoring the units of the period and angular frequency is easily done by appropriately rescaling by the magnitude of the decay vector $\mathbf{\Gamma}$:
\begin{equation}
    {\rm P} \: = \: \frac{\pi}{|\mathbf{E}|\sqrt{1-r^2}} \ , \qquad \omega \: = \: 2|\mathbf{E}|\,\sqrt{1-r^2}\ .
\end{equation}
Notice that as $r$ approaches $1$, these expressions imply an oscillation period ${\rm P}$ that tends to infinity. Thus, in the extreme case $r=1$,  the co-decaying Bloch vector $\mathbf{b}$ of CUQ no longer oscillates, independently of any initial condition for $\mathbf{b}(0)$ at $t=0$. This non-oscillatory feature must be anticipated for such extremal CUQs, since the two energy eigenstates of the two-level quantum system become exactly degenerate in this case. In~\cite{Pilaftsis:1997dr}, it was shown that such critical non-oscillatory two-level quantum systems correspond to  exceptional forms of the effective Hamiltonian which takes on the Jordan form. In particular, such extremal CUQs were found to maximise the phenomenon of resonant CP~violation through mixing of states.

\subsection{Coherence--Decoherence Oscillations}

As another important application of the analysis presented in this section, let us consider a critical unstable qubit which is initially prepared in a fully mixed state with a vanishing Bloch vector, i.e.~$\vecb(0) = \mathbf{0}$, at $t=\tau=0$. 

This change of initial condition greatly alters the methods required to find analytic solutions to the master evolution equation. As can be seen in appendix~\ref{app:Mixed}, where taking $|\vecb|=1$ leads to divergences. However, following the argument presented there, one can find the mixed state solution, and show that the magnitude of $\vecb(\tau)$ follows:
\begin{equation}
    |{\bf b}(\tau)|^2\ =\ 1\: -\:  (1-r^2)^2\,\lrsb{1-r^2 \, \cos \lrb{\dfrac{\sqrt{1-r^2}}{r} \,\tau} }^{-2}   \;.
    \label{eq:b_crit}
\end{equation}
Notice that for $r<1$, the magnitude of the Bloch vector, $|\vecb(\tau )|$, will oscillate between zero and a maximum value given by ${\rm max}(\absb) = 1 - (1-r^2)^2/(1+r^2)^2$, while the period of oscillation will be the same as the pure state, as per the equivalence of CUQs. Although, we should comment that in the limiting case $r = 1$, we have 
\begin{equation}
|{\bf b}(\tau)|^2\: =\: 1\,-\,\frac{4}{(2+\tau^2)^2}\;,
\end{equation}
so the critical unstable qubit will tend asymptotically to a pure state in the co-decaying frame, as~${\tau \to \infty}$.

\begin{figure}
    \centering
    \includegraphics[width=0.8\linewidth]{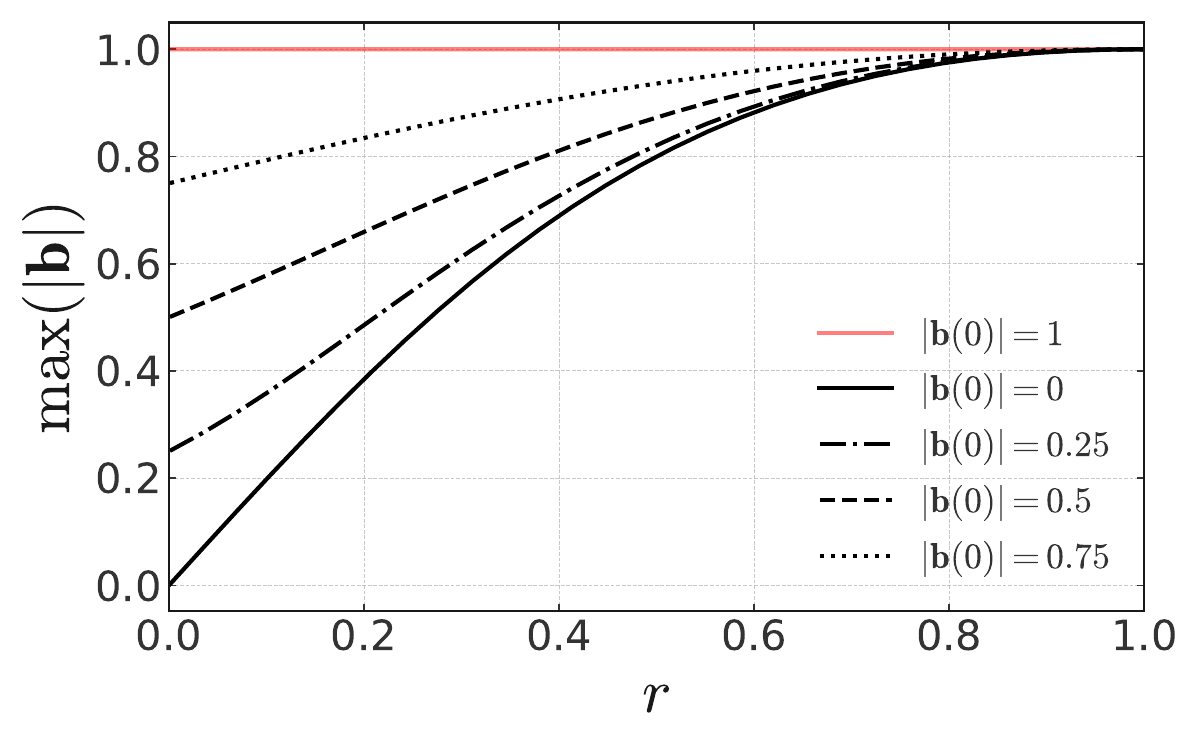}
    \caption{The maximum value of the magnitude of the co-decaying Bloch vector $\mathbf{b}(\tau )$ as a function of the parameter~$r$, for a critical unstable qubit. A set of different initial conditions for $\mathbf{b}(0)\parallel \boldsymbol{\gamma}$ was assumed.}
    \label{fig:bMax}
\end{figure}

Figure~\ref{fig:bMax} shows the maximum values of~$|\mathbf{b}(\tau )|$ as a function of~$r$, for a critical unstable qubit,
while assuming a set of representative initial conditions for~$|\mathbf{b}(0)|$
and taking $\mathbf{b}(0)\parallel \boldsymbol{\gamma}$.
Most remarkably, we find that critical unstable qubit systems will display {\em coherence--decoherence} oscillations in time when analysed in its co-decaying frame. This is in stark contrast to the dynamics of a stable qubit, where no such phenomena are possible.

\section{Fourier Coefficients of CUQs}

Having stated the basic equation governing the time evolution of the co-decaying Bloch vector, $\mathbf{b}(\tau)$, in the previous section [cf.~\eqref{eq:AngSol}], we will now consider observables that would allow us to physically distinguish CUQs from ordinary unstable qubits. In general, observable quantities are often proportional to the projection of ${\bf b}(\tau)$ along a particular axis, e.g.~along the unit decay vector~$\boldsymbol{\gamma}$. Although the expression in equation~\eqref{eq:AngSol} could be used to fit a signal and extract estimates for $r$, it may still be difficult to reach a precision high enough to determine these values if $r \ll 1$. However, as discussed in the previous section, CUQs have the unique feature that their oscillation is \textit{anharmonic} rather than harmonic. Therefore, we propose searching for Fourier modes beyond the principal harmonic mode as an alternative and more efficient method of extracting the key model parameter $r$. With this in mind, we consider in the following two subsections a Fourier series decomposition of the expected signal arising from~a CUQ.

\subsection{Evaluation of Fourier Coefficients}

As a reference model, we assume a system which is initially prepared so that $\be(0) = \mathbf{e}\times \boldsymbol{\gamma}$. Additionally, since $\widehat{\omega}\tau = \omega t$ is a dimensionless quantity, we choose to evaluate the Fourier coefficients in terms of the dimensionless quantities, $r$, $\widehat{\omega}$, and~$\tau$. By choosing to study the evolution of $\vecb(\tau)$ with this initial condition, we are able to fully exploit the symmetries present in the evolution and express $\be \cdot \boldsymbol{\gamma}$ as an exclusively odd-term Fourier series:
\begin{equation}
  \label{eq:bdotgamma}
    \be (\tau) \cdot \boldsymbol{\gamma} \: = \: \sum_{n=1}^\infty \, c_n \, \sin \lrb{n\widehat{\omega} \tau} \, , 
\end{equation}
with
\begin{equation}\label{eq:FourierIntegral}
    c_n \: = \: -\frac{2}{\widehat{\rm P}} \int_{-\widehat{\rm P}/2}^{\widehat{\rm P}/2} \, \sin(n\widehat{\omega} \tau) \, \sin \theta \: \de \tau \, .
\end{equation}
By making use of~\eqref{eq:AngSol}, it is not difficult to find an expression for $\be \cdot \boldsymbol{\gamma} \, = \, - \sin \theta$, in terms of~$\tau$, by inverting the tangent half-angle transformations, 
\begin{subequations}
    \begin{align}
    \sin \theta \: &= \: \frac{2 \tan \frac{\theta}{2}}{1+\tan^2\frac{\theta}{2}} \: = \: -\sqrt{1-r^2}\frac{\sin \lrb{\widehat{\omega} \tau}}{1-r\cos\lrb{\widehat{\omega}\tau}} \ , \label{eq:sinEvol}\\
    \cos \theta \: &= \: \frac{1-\tan^2 \frac{\theta}{2}}{1+\tan^2 \frac{\theta}{2}} \: = \: \frac{\cos \lrb{\widehat{\omega}\tau} - r}{1 - r\cos\lrb{\widehat{\omega} \tau}} \, , \label{eq:cosEvol}
\end{align}
\end{subequations}
where the last formula for $\cos\theta$ was included for completeness. After inserting the expression for $\sin\theta$ in~\eqref{eq:sinEvol} into~\eqref{eq:FourierIntegral}, the Fourier coefficients may be computed as follows:
\begin{equation}\label{eq:cn}
    c_n \: = \: \frac{2\sqrt{1-r^2}}{\widehat{\rm P}} \int_{-\widehat{\rm P}/2}^{\widehat{\rm P}/2} \, \sin(n\widehat{\omega} \tau) \, \frac{\sin \lrb{\widehat{\omega} \tau}}{1-r\cos\lrb{\widehat{\omega}\tau}} \, \de \tau \, .
\end{equation}

After making use of careful substitutions, this integral can be evaluated using the residue theorem, with consideration to only include the relevant poles. In particular, only one of the two poles present will contribute to the Fourier integral, related to the application of the condition: $0<r<1$. Taking this fact into account, the Fourier series coefficients are found to be
\begin{align}\label{eq:FourierCoeffC}
    c_n \: = \: 2 \frac{\sqrt{1-r^2}}{r}\,\bigg(\frac{1}{r}\,-\,\frac{\sqrt{1-r^2}}{r}\bigg)^n \, .
\end{align}
We should clarify here that there is no divergence 
in $c_n$ for very small values of $r$. To see this, we utilise
the MacLaurin series expansion for the expression of interest,
\begin{equation}
    \frac{1}{r}-\frac{\sqrt{1-r^2}}{r} \ \simeq \ \frac{1}{r} - \bigg(\frac{1}{r} - \frac{r}{2} + \mathcal{O}(r^3)\bigg) \ \simeq \  \frac{r}{2} + \mathcal{O}(r^3)\, .
\end{equation}
As can be seen in the above, the offending $1/r$ terms cancel  as $r\to 0^+$, thereby rendering the expression for $c_n$ finite. Thus, we find that the leading term of $c_n$ is of order $r^{n-1}$. 
In fact, in the limit $r\to 0^+$, we recover 
the typical Rabi oscillations for which only the principal harmonic survives, i.e.
\begin{equation}
    \lim_{r\to 0^+} \, c_1\: =\: 1 \, , \qquad \lim_{r\to 0} \, c_{n \, \geq \, 2}\: =\: 0 \, .
\end{equation}

To evaluate the Fourier coefficients for the projection of $\be (\tau )$ along $\mathbf{e}\times\boldsymbol{\gamma}$, we follow an identical approach. In this case, we use the expression for $\cos\theta$ given in~\eqref{eq:cosEvol} and expand over even Fourier modes as follows:
\begin{equation}
  \label{eq:bexgamma}
    \be (\tau)\cdot(\mathbf{e}\times\boldsymbol{\gamma}) \: = \: d_0\, +\, \sum_{n=1}^\infty \, d_n \cos(n\widehat{\omega} \tau) \, ,
\end{equation}
In general, one finds that despite different analytic forms of their integrals, the Fourier coefficients of the even series are identical to those found in the odd series, i.e.~$d_n = c_n$; however, in the case of the zeroth component of the even Fourier series, one should again take care to only include the relevant poles.

\begin{figure}[t!]
    \centering
    \begin{subfigure}{0.49\linewidth}
    \centering
    \includegraphics[width=\linewidth]{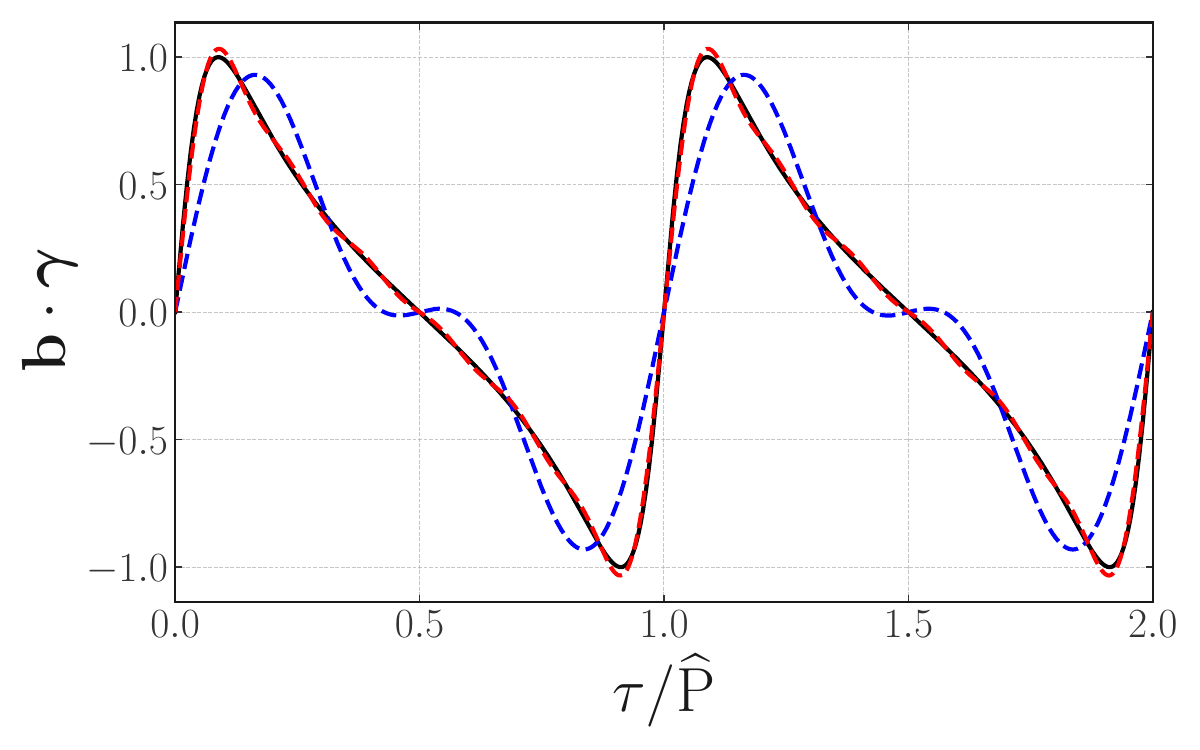}
    \caption{\empty}
    \label{fig:FourierE}
    \end{subfigure}
    \begin{subfigure}{0.49\linewidth}
    \centering
    \includegraphics[width=\linewidth]{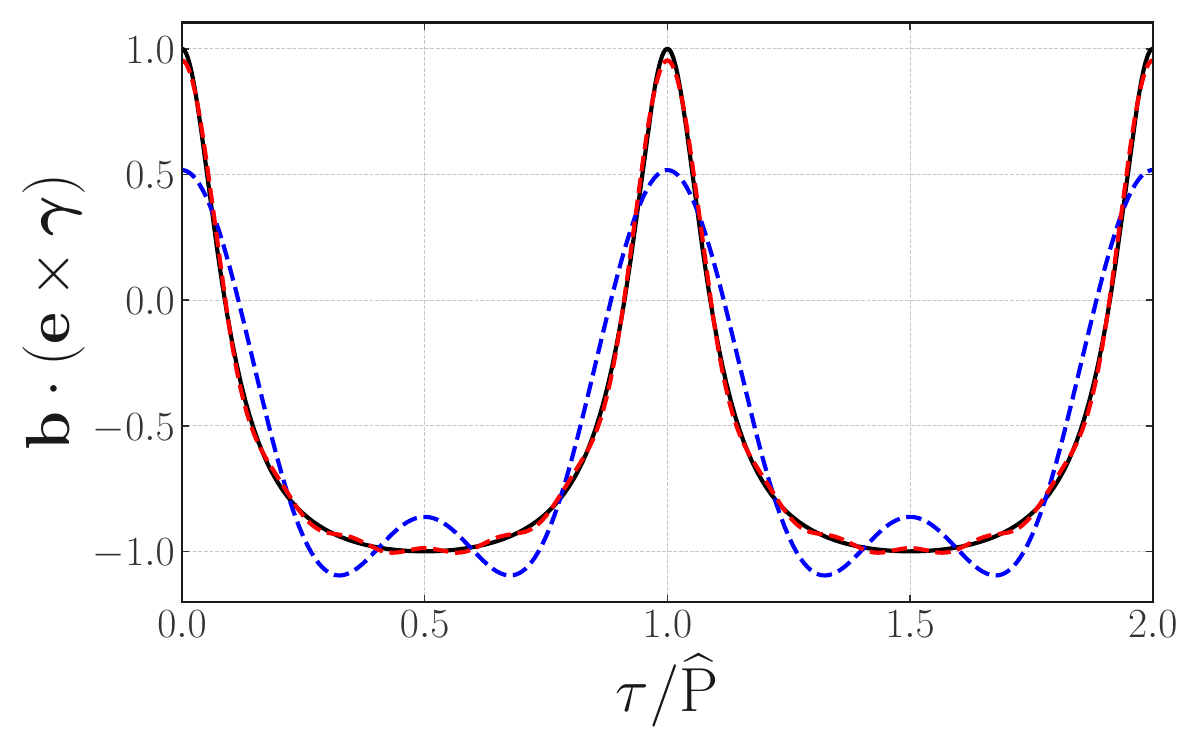}
    \caption{\empty}
    \label{fig:FourierEGam}
    \end{subfigure}
    \caption{Fourier series approximations for a CUQ with $r=0.85$ and $\be(0)=\mathbf{e}\times\boldsymbol{\gamma}$. The black line is the exact evolution of ${\bf b}(\tau)$, the blue dashed line is the Fourier series including terms up to and including $n=2$, and the red dashed line is the Fourier series including terms up to and including $n=6$. The left panel ($a$) shows the projection onto $\boldsymbol{\gamma}$, and the right panel ($b$) shows the projection onto $\mathbf{e}\times\boldsymbol{\gamma}$.}
    \label{fig:Fourier_Proj}
\end{figure}

In Figure~\ref{fig:Fourier_Proj}, we show only the projections of ${\bf b}(\tau )$ along the directions $\boldsymbol{\gamma}$ and $\mathbf{e}\times\boldsymbol{\gamma}$. According to our reference model, we~assume the initial condition $\vecb(0) = \mathbf{e}\times\boldsymbol{\gamma}$ and also take $r=0.85$. The black lines show the exact evolution of CUQ oscillations, the blue dashed lines show the Fourier series up to and including the $n=2$ Fourier coefficient, and the red dashed lines show the Fourier series up to and including the $n=6$ Fourier coefficient. As shown in~\cite{Karamitros:2022oew}, CUQs exhibit an inhomogeneous rotation of ${\bf b}(\tau )$, which slows down and freezes as $r$ approaches~1. Moreover, in Figure~\ref{fig:Fourier_Proj}, we observe that the oscillation profile of ${\bf b}(\tau )$ strongly depends on the particular axis onto which $\vecb (\tau )$ is projected, as could happen in realistic experiments. In~particular, we see that projections along $\boldsymbol{\gamma}$ have a triangular wave form, whereas the projection along $\mathbf{e}\times\boldsymbol{\gamma}$ features sharp peaks and wide troughs. Nonetheless, we have verified that the two different oscillation profiles displayed in the two panels of Figure~\ref{fig:Fourier_Proj} preserve the condition expected from a pure qubit, $|\vecb(\tau)|=1$, throughout their evolution.

With the aid of the analytic results derived in this section, we can approximate the odd and even Fourier series projections of~$\be (\tau)$ given in~\eqref{eq:bdotgamma} and~\eqref{eq:bexgamma} up to any given order $N$, by including all Fourier modes $n \leq N$. In this respect, it is worth noting that the Fourier coefficients, $c_n$ and $d_n$, follow a sequence of geometric progression. Based on this property, we~define the {\em anharmonicity factors},
\begin{equation}
    \mathcal{C}_n \: \equiv \: \frac{c_{n+1}}{c_n}\, , \qquad  \mathcal{D}_{n-1} \: \equiv \: \frac{d_{n}}{d_{n-1}} \, ,
\end{equation}
with $n\geq 1$. These factors are independent of~$n$, and as such, they can be regarded as physical observables, as their measurements allow us to determine the key model parameter~$r$ from oscillation data.
For instance,  for the odd Fourier series given in~\eqref{eq:bdotgamma}, we find 
\begin{equation}
    r \: = \: \frac{2\, \mathcal{C}_n}{\mathcal{C}_n^2+1}\: \overset{\mathcal{C}_n\ll1}{\approx}\: 2\, \mathcal{C}_n\, +\, \mathcal{O}(\mathcal{C}_n^{3})\, ,
\end{equation}
for all $n\geq 1$. By analogy, for the even Fourier series in~\eqref{eq:bexgamma}, we may deduce, independently of~$n$, the value of $r$ from the the following relations:
\begin{subequations}
\begin{align}
     \label{eq:Fourier_Ratios}
    r \: &= \: \frac{1}{\sqrt{1+\frac{1}{4}\mathcal{D}_0^2}} \: \overset{\mathcal{D}_0\gg 1}{\approx} \: \frac{2}{\mathcal{D}_0} + \mathcal{O}(\mathcal{D}_0^{-3})\, ,\\ r \: &= \: \frac{2\, \mathcal{D}_n}{\mathcal{D}_n^2+1} \: \overset{\mathcal{D}_n\ll1}{\approx} \: 2\, \mathcal{D}_n + \mathcal{O}(\mathcal{D}_n^{3})\,, \qquad \mbox{for $n\geq 1$.}
\end{align}
\end{subequations}
A significant benefit of estimating $r$ by taking ratios of the Fourier coefficients is that some of the amplitude dependence drops out. However, there are some non-trivial aspects to be considered. For pure state oscillations where the projection of ${\bf b}(\tau)$ has an amplitude $R<1$, the actual value of $r$ extracted from the anharmonicity factors will be an effective value, which is denoted as $\tilde{r}$. In order to map the latter onto the true value of $r$, we employ the relation,
\begin{equation}\label{eq:r_effective}
    r \: = \: \frac{\tilde{r}}{\sqrt{R^2 + \tilde{r}^2 \lrb{1-R^2}}}\ . 
\end{equation}
Note that for $\tilde{r}\ll 1$, the value $r$ is obtained by the simple rescaling: $r= \tilde{r}/R\: +\: \mathcal{O}(\tilde{r}^3)$.  

The origin of the correction to $r$ given in equation~\eqref{eq:r_effective} is rooted in the geometry of general CUQs. It~is the fact that $\be (\tau)$ is not really oscillating about the energy vector, $\mathbf{e}$, but truly about a linear combination of $\mathbf{e}$ and $\boldsymbol{\gamma}\times\be(0)$.


\section{Application to $B^0\bar{B}^0$ Oscillations}\label{sec:Bmeson}

In this section, we apply the approximate Fourier series approach developed in the previous section to study anharmonicities in the oscillation profile of a meson–antimeson system. For definiteness, we will focus on the $B^0\bar{B}^0$-meson system. 

\subsection{Bloch Sphere Parameters for the {\em B}--Meson System}

To implement our approach more efficiently, we must first translate the relevant $B$-meson observables, which are derived from the effective Hamiltonian and were extensively discussed in the literature~\cite{Kabir:1989dd,Grimus:1988,Paschos:1989ur,Buras:1990fn,Neubert:1993mb,Nierste:2004uz}, into the Bloch sphere parameters introduced in Section~\ref{sec:Bloch}. In terms of these parameters, the mass and width differences of a $B$-meson system are given by
\begin{equation}
   \label{eq:DEDGamma}
    \Delta E \: = \: 2|\mathbf{E}| \Re {\sqrt{1-r^2 -2ir\cos\theta_{\mathbf{e}\boldsymbol{\gamma}}}} \, , \qquad \Delta \Gamma \: = \: -4|\mathbf{E}| \Im{\sqrt{1-r^2 -2ir\cos\theta_{\mathbf{e}\boldsymbol{\gamma}}}} \, .
\end{equation}
Including the CP-violation measure, $|q/p|$, allows one to uniquely determine the oscillation parameters\- of the effective Hamiltonian~${\rm H}_{\rm eff}$ by the relation~\cite{Karamitros:2022oew},
\begin{equation}
   \label{eq:qoverp4}
    \left| \frac{q}{p} \right|^4 \: = \: \frac{1+r^2 - 2r \sin \theta_{\mathbf{e}\boldsymbol{\gamma}}}{1+r^2 + 2r \sin \theta_{\mathbf{e}\boldsymbol{\gamma}}} \ .
\end{equation}
From~\eqref{eq:qoverp4}, it is then easy to see that when either $r = 0$ or $\sin \theta_{\mathbf{e}\boldsymbol{\gamma}} = 0$, one has $|q/p| =1$, so~CP violation through mixing of states will vanish. In this case, ${\rm H}_{\rm eff}$ may be diagonalised by means of unitary transformations, and dispersive and dissipative effects become distinct from one another. In contrast, for a CUQ, for which~$\cos\theta_{\mathbf{e}\boldsymbol{\gamma}}=0$ and~$r<1$, CP violation induced by the mixing of the two quantum states, which manifests itself in a non-zero value of $|(|q/p| -1)|$, is maximized at any given $r$. Moreover, for a CUQ, the expressions given in~\eqref{eq:DEDGamma} and~\eqref{eq:qoverp4} simplify to
\begin{equation}
    \Delta E \: = \: 2|\mathbf{E}|\sqrt{1-r^2} \, , \qquad \Delta \Gamma \: = \: 0 \, , \qquad \left| \frac{q}{p} \right| \: = \: \sqrt{\frac{1 \pm r}{1 \mp r}} \; ,
\end{equation}
where the sign of $\theta_{\mathbf{e}\boldsymbol{\gamma}}$ is determined by the signs contained within the expression for $|q/p|$. 

\begin{table}[t!]
    \centering
    \begin{tabular}{|c||c|c|c|}
        \hline
        {} & $\Delta E \, [{\rm ps}^{-1}]$ &  $\Delta\Gamma \, [{\rm ps}^{-1}]$  & $|q/p|-1$\\[0.1cm]
        \hline\hline
        Experiment&  $0.5069 \pm 0.0019$ & $(0.7 \pm 7)\times 10^{-3}$ & $ (1.0\pm 0.8) \times 10^{-3} $\\
        \hline
        Theory &  $0.535\pm 0.021$ & $\lrb{2.7 \pm 0.4}\times 10^{-3}$ & $(2.6 \pm 0.3) \times 10^{-4}$\\[0.1cm]
        \hline\hline
        {} & $r$ &  $\theta_{\mathbf{e}\boldsymbol{\gamma}} [^\circ]$  & $|\mathbf{E}| [\textrm{ps}^{-1}]$\\[0.1cm]
        \hline\hline
        Experiment&  $(1\pm 4) \times 10^{-3}$ & $-90 \pm 90$ & $ 0.253 \pm 0.001 $\\
        \hline
        Theory &  $(2.5 \pm 0.4) \times 10^{-3}$ & $-5 \pm 3$ & $0.28 \pm 0.01$\\[0.1cm]
        \hline
    \end{tabular}
    \caption{\em 
    The experimental values for the various CP-violating parameters are taken from~\cite{HeavyFlavorAveragingGroupHFLAV:2024ctg} and the most recent theoretical SM predictions from are extracted from~\cite{Albrecht:2024oyn}.
    Also shown are the model parameters, $r$, $\theta_{\mathbf{e}\boldsymbol{\gamma}}$ and~$|\mathbf{E}|$, as deduced from the $B^0$-meson oscillation data, along with their SM predictions.}
    \label{tab:MesonData}
\end{table}

In Table~\ref{tab:MesonData} we show the central values and bounds at the $1\,\sigma$ Confidence Level (CL) for the frequently quoted quantities $\Delta E$, $\Delta\Gamma$ and $|q/p| -1$, as well as the respective values for the relevant Bloch sphere parameters $r$, $\theta_{\mathbf{e}\boldsymbol{\gamma}}$ and $|{\bf E}|$, as these are estimated from both experimental and theoretical analyses. For definiteness, the experimental values are taken from~\cite{HeavyFlavorAveragingGroupHFLAV:2024ctg} and the most recent theoretical predictions in the Standard Model (SM) can be found in~\cite{Albrecht:2024oyn}. In the SM, the $B^0\bar{B}^0$ system is predicted to be far away from a CUQ realisation. Nevertheless, physics beyond the SM could be responsible for generating the CUQ dynamics in the $B^0_{\rm d}$-meson system.

\subsection{CP-Violating Observables}

In general, the oscillation of neutral mesons, such as the $K^0\bar{K}^0$- and $B^0\bar{B}^0$-meson systems, is theoretically quantified through some observable of flavour asymmetry~\cite{Kabir:1989dd}, e.g.
\begin{equation}
    \delta(t) \: = \: \frac{N_s(t) - \overline{N}_s(t)}{N_s(t) + \overline{N}_s(t)} 
    \, ,
    \label{eq:delta_def}
\end{equation}
where $N_s$ is the number of particles of the signal in the initial state, and $\overline{N}_s$ the number of particles in the CP-conjugate state. 
This result can be expressed in terms of the co-decaying Bloch vector as follows:
\begin{equation}
    \delta(t)  \: = \: \frac{\mathbb{P}({\rm P}^0) - \mathbb{P}(\bar{\rm P}^0)}{\mathbb{P}({\rm P}^0) + \mathbb{P}(\bar{\rm P}^0)}  \: = \: \frac{{\rm Tr}{\sigma^3 \rho(t)}}{{\rm Tr}\, \rho(t)} \: = \: {\rm b}_3(t) \, ,
    \label{eq:delta_b3}
\end{equation}
with ${\rm P}^0 = K^0,\, B^0, D^0$.
In the CP basis, assuming CPT invariance of the effective Hamiltonian leads to the constraints~\cite{Grimus:1988,Kabir_1996,
ParticleDataGroup:2024cfk}
\begin{equation}
    {\rm E}_{11} \: = \: {\rm E}_{22}\,,\qquad 
    \Gamma_{11} \: = \: \Gamma_{22}\, ,    
\end{equation}
which implies ${\rm E}_3=\Gamma_3=0$. For CUQs (where $\mathbf{e}\cdot\boldsymbol{\gamma}=0$), the last constraint is translated to  
\begin{equation}
    \mathbf{e}\times\boldsymbol{\gamma} \: = \: \begin{pmatrix}
        0,  \, 0, \, 1 
    \end{pmatrix} \, .
\end{equation}
Therefore, with the help of~\eqref{eq:delta_b3}, the flavour asymmetry in~\eqref{eq:delta_def} takes on the simple form,
\begin{equation}
    \delta(t)  \: = \: {\rm b}_3(t) \: = \: \vecb(t) \cdot (\mathbf{e}\times\boldsymbol{\gamma})\, .
    \label{eq:delta_b_proj}
\end{equation}

\begin{figure}[t!]
    \centering
    \begin{subfigure}{0.49\linewidth}
    \centering
    \includegraphics[width=\linewidth]{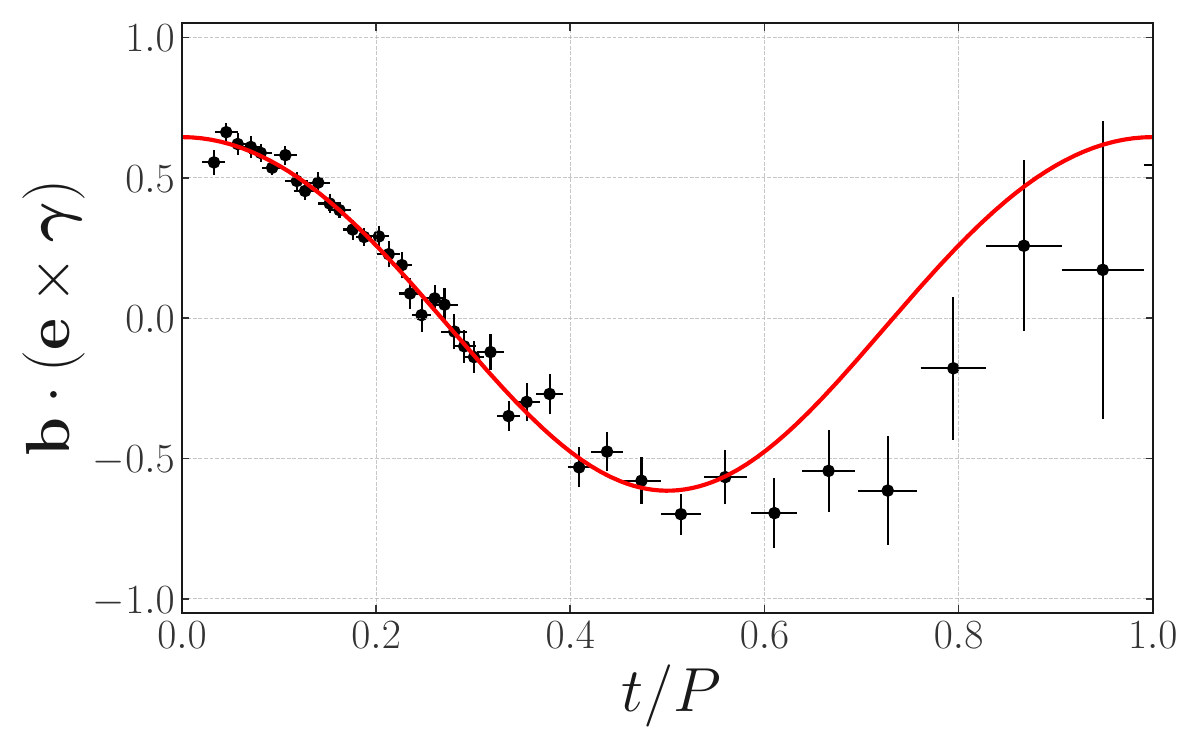}
    \caption{\empty}
    \label{fig:B2D2011}
\end{subfigure}
\begin{subfigure}{0.49\linewidth}
    \centering
    \includegraphics[width=\linewidth]{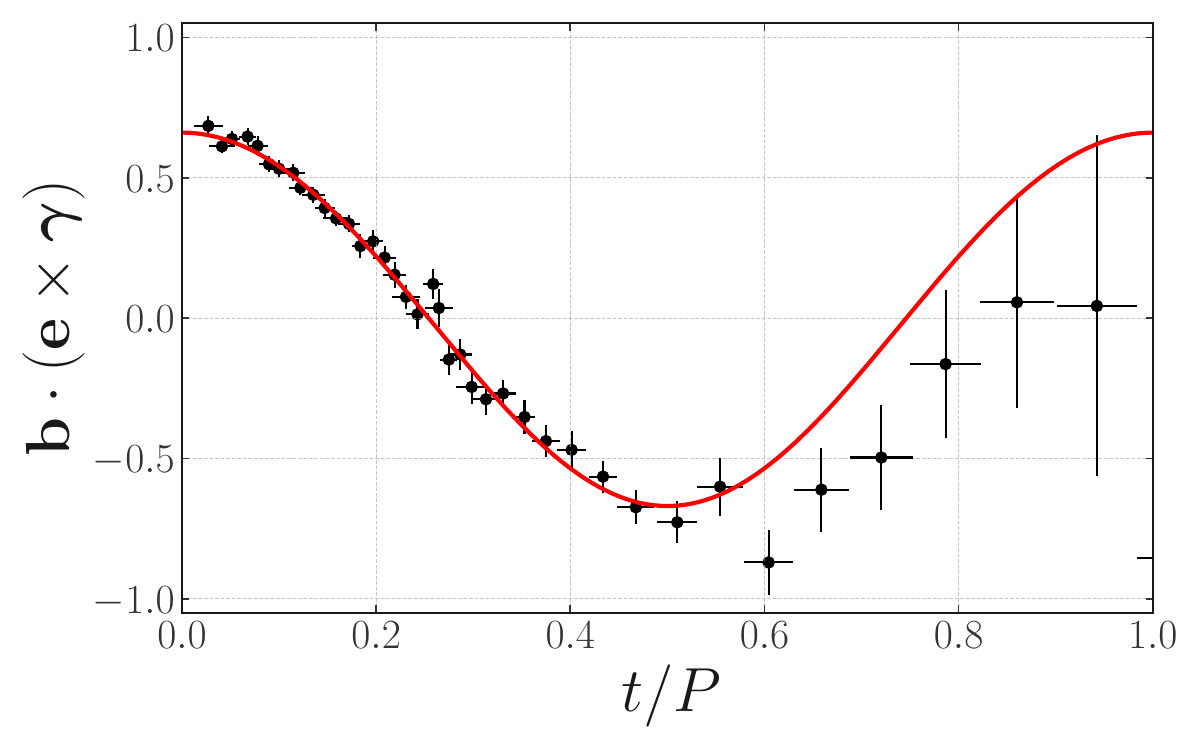}
    \caption{\empty}
    \label{fig:B2DStar2011}
\end{subfigure}
\begin{subfigure}{0.49\linewidth}
    \centering
    \includegraphics[width=\linewidth]{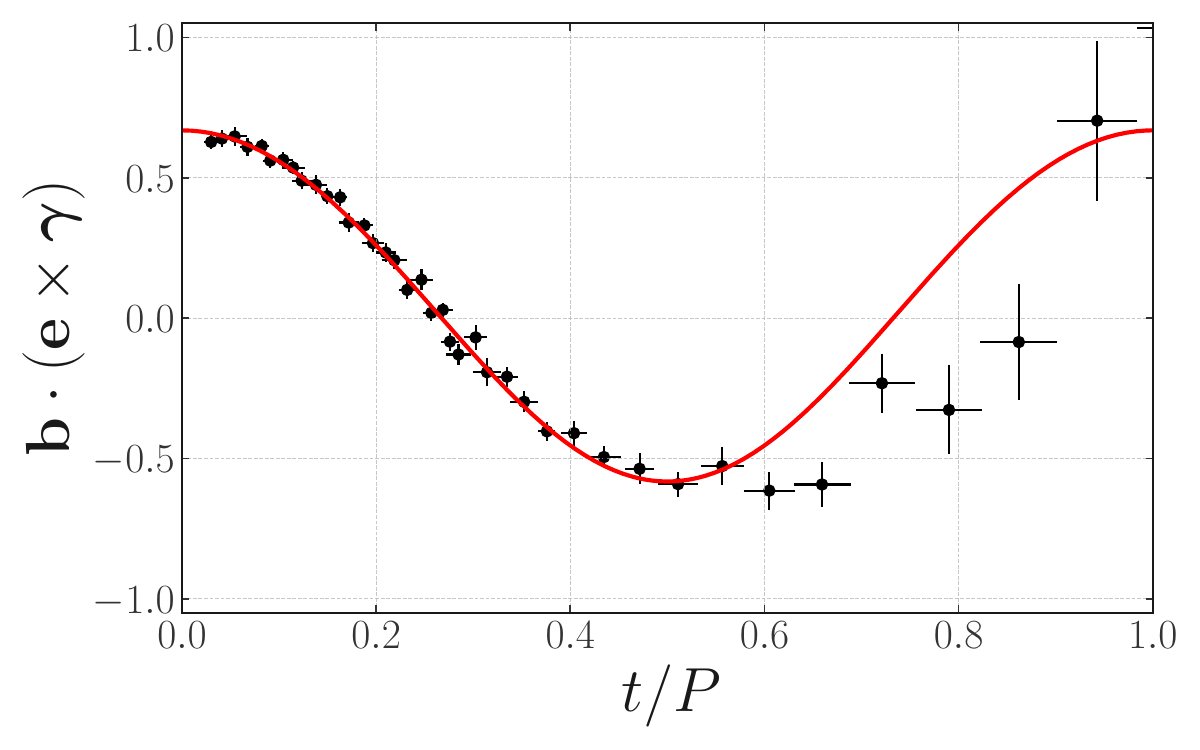}
    \caption{\empty}
    \label{fig:B2D2012}
    \end{subfigure}
    \begin{subfigure}{0.49\linewidth}
    \centering
    \includegraphics[width=\linewidth]{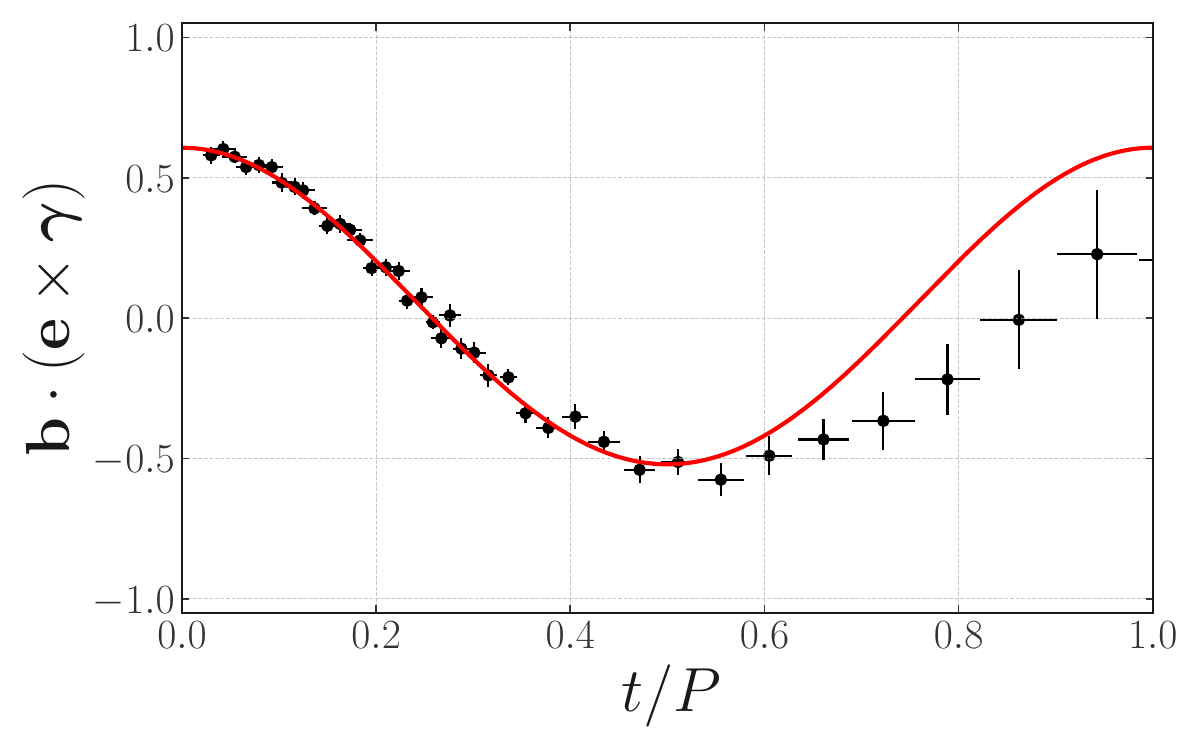}
    \caption{\empty}
    \label{fig:B2DStar2012}
    \end{subfigure}
    \caption{Fourier series fit for the data given in~\cite{LHCb:2016gsk}, for the year 2011: panels ($a$) and ($b$), and the year 2012: panels ($c$) and ($d$). Panels ($a$) and ($c$) show the flavour asymmetry for the channel $B^0_{\rm d}\to D^-\mu^+\nu_\mu X$ and panels ($b$) and ($d$) for the channel $B^0_{\rm d}\to D^{-*}\mu^+\nu_\mu X$. The black data points and their associated error bars have been rewritten from their original form into projections of  along the $\mathbf{e}\times\boldsymbol{\gamma}$ direction. For this fit, we make use of the first two harmonics of the Fourier series, as well as the constant term.}
    \label{fig:Fourier_B_meson}
\end{figure}

Restricting our analysis to the $B^0\bar{B}^0$-meson system, we can extract $\vecb(t)\cdot\lrb{\mathbf{e}\times\boldsymbol{\gamma}}$ and use this information to evaluate the Fourier coefficients of the oscillation profile. The original analysis of LHCb~\cite{LHCb:2016gsk} identifies the mass splitting of the $B$-meson system, $\Delta m_{\rm d}$, with the measured angular frequency~$\omega$ of the mixing asymmetry oscillation. Consequently, we equate $\omega$ with $\Delta m_{\rm d}$. The Fourier coefficients are calculated by making use of a multiple linear regression of the projected Bloch vector, $\vecb(t)\cdot\lrb{\mathbf{e}\times\boldsymbol{\gamma}}$, against the Fourier modes, $\cos(n \omega t)$, where the Fourier coefficients are the regression values that minimize the $\chi^2$ of the fit.

Figure~\ref{fig:Fourier_B_meson} shows the mixing asymmetry oscillation data as a projection of $\vecb (t)$ onto ${\mathbf{e}\times\boldsymbol{\gamma}}$. Figures~\ref{fig:B2D2011} and~\ref{fig:B2DStar2011} show the $2011$ data for the decay processes, $B^0_{\rm d} \to D^-\mu^+\nu_\mu X$ and ${B^0_{\rm d} \to D^{-*}\mu^+\nu_\mu X}$, respectively. Likewise, Figures~\ref{fig:B2D2012} and~\ref{fig:B2DStar2012} exhibit the corresponding $2012$ data for the above two decays.

The red line is a truncated Fourier series fit by taking into account the sum of $n=0,\, 1$ and $2$ harmonics. The Fourier coefficients extracted from this fit, with their respective errors, are given in Table~\ref{tab:Fourier_Coeffs}. Alongside these Fourier coefficients, we check whether each coefficient deviates significantly from zero, where the latter value represents our {\em null hypothesis}. The $p$-value of each coefficient is also stated in Table~\ref{tab:Fourier_Coeffs}, which has been calculated using a standard $t$-distribution with mean $\mu=0$. We have found small $p$-values for the Fourier coefficient, $d_1$, which means the hypothesis of the basic sinusoidal harmonic becomes statistically significant.  

\begin{table}[t!]
    \centering
    \begin{tabular}{|c||c|c|c|c|}
        \hline
         - & \multicolumn{2}{c|}{$B_{\rm d}^0 \to D^-\mu^+\nu_\mu X \, (2011)$} & \multicolumn{2}{c|}{$B_{\rm d}^0 \to D^{-*}\mu^+\nu_\mu X\, (2011)$}  \\[0.1cm]
        \hline\hline
        {} & $d_n \pm \delta d_n$ & $p$-value & $d_n \pm \delta d_n$ & $p$-value \\ 
        \hline
        $d_0$ & $0.04 \pm 0.12$ & $74\%$ & $0.006 \pm 0.137$ & $96\%$ \\
        \hline
        $d_1$ & $0.630 \pm 0.007$ & $< 0.01\%$ & $0.665 \pm 0.007$ & $<0.01\%$ \\
        \hline
        $d_2$ & $-0.03 \pm 0.01$ & $ 6\%$ & $0.01 \pm -0.01$ & $42\%$ \\
        \hline\hline
        - & \multicolumn{2}{c|}{$B_{\rm d}^0 \to D^-\mu^+\nu_\mu X \, (2012)$} & \multicolumn{2}{c|}{$B_{\rm d}^0 \to D^{-*}\mu^+\nu_\mu X\, (2012)$}  \\[0.1cm]
        \hline\hline
        {} & $d_n \pm \delta d_n$ & $p$-value & $d_n \pm \delta d_n$ & $p$-value \\ 
        \hline
        $d_0$ & $0.06 \pm 0.07 $ & $45\%$ & $0.04 \pm 0.06$ & $58\%$\\
        \hline
        $d_1$ & $ 0.625 \pm 0.004$ & $<0.01\%$ & $0.564 \pm 0.003$ & $<0.01\%$ \\
        \hline
        $d_2$ & $ -0.013 \pm 0.009 $ & $18\%$ & $0.007 \pm 0.008$ & $37\%$ \\
        \hline
    \end{tabular}
    \caption{\em Fourier coefficients extracted from the regression fit of the data given in~\cite{LHCb:2016gsk}. The errors are estimated as a combination of the errors associated with individual data points as well as the errors inherent to fitting the Fourier coefficients. For completeness, we additionally give p-values to provide an assessment of the significance of each Fourier coefficient as a departure from zero.}
    \label{tab:Fourier_Coeffs}
\end{table}

\begin{table}[t!]
    \centering
    \begin{tabular}{|c||c|c|c|c|}
        \hline
         {} & \multicolumn{2}{c|}{$B_{\rm d}^0 \to D^-\mu^+\nu_\mu X \, (2011)$} & \multicolumn{2}{c|}{$B_{\rm d}^0 \to D^{-*}\mu^+\nu_\mu X\, (2011)$}  \\[0.1cm]
        \hline\hline
        {} & $ \mathcal{D}_n $ & $r$ & $\mathcal{D}_n $ & $r$ \\
        \hline
        $\mathcal{D}_0$ & $15 \pm 44$ & $0.2\pm 0.6$ & $109 \pm 2464$ & $0.03 \pm 0.06$ \\
        \hline
        $\mathcal{D}_1$ & $0.04 \pm 0.02$ & $0.13 \pm 0.06$ & $0.02 \pm 0.02$ & $0.05\pm0.06$ \\
        \hline\hline
        {} & \multicolumn{2}{c|}{$B_{\rm d}^0 \to D^-\mu^+\nu_\mu X \, (2012)$} & \multicolumn{2}{c|}{$B_{\rm d}^0 \to D^{-*}\mu^+\nu_\mu X\, (2012)$}  \\[0.1cm]
        \hline\hline
        {} & $ \mathcal{D}_n $ & $r$ & $\mathcal{D}_n $ & $r$ \\ 
        \hline
        $\mathcal{D}_0$ & $11 \pm 15 $ & $0.3 \pm 0.3$ & $16 \pm 29$ & $0.2 \pm 0.4$\\
        \hline
        $\mathcal{D}_1$ & $ 0.02 \pm 0.01$ & $0.06 \pm 0.05$ & $0.01 \pm 0.02$ & $0.05 \pm 0.06$ \\
        \hline
    \end{tabular}
    \caption{\em Estimations of the anharmonicity factors and values of $r$ for the $2011$ and $2012$ data points provided in~\cite{LHCb:2016gsk}. Alongside these central values, we provide estimates of their associated $1\sigma$ error bounds.}
    \label{tab:r_estimates}
\end{table}

As can also be seen in Table~\ref{tab:Fourier_Coeffs}, the other two harmonics, $d_0$ and $d_2$, acquire sizeable $p$-values through our analysis, which point to low statistical significance from the null hypothesis. As~a result, it may be difficult to reach a high quality estimate for the dimensionless parameter~$r$. In~the following subsection, we will explain the reasons for this low significance. As stated in~\cite{LHCb:2012mhu,LHCb:2016gsk}, it is worth drawing attention to the fact that the lifetime of $B^0$ mesons is about $6~{\rm ps}$, whereas the oscillation period is measured to be around $12~{\rm ps}$. Hence, a significant number of $B^0$ mesons have decayed before half an oscillation has been observed. As a consequence, data points after $t\simeq \textrm{P}/2$ are of lower precision, and therefore the error bars widen with time, as this is reflected in Figure~\ref{fig:Fourier_B_meson}.

For $N$ Fourier coefficients, $N-1$ estimates for $r$ can be obtained. These results are presented in Table~\ref{tab:r_estimates}. As far as the Fourier coefficients themselves are concerned, the large errors pervade into the estimates of the anharmonicity factors, ${\cal D}_0$ and ${\cal D}_1$, and the model parameter~$r$. From the data analysed here, we can take a weighted average of the values for~$r$ giving the central estimate $r = 0.07 \pm 0.03$. We see that given the present data set and the current analysis, it is difficult to attain an estimate of $r$ which differs significantly from~zero. The estimates we have shown in~Table~\ref{tab:r_estimates} deviate from the Rabi oscillation assumption with very low statistical significance amounting to values of~$p=8\%$. As we clarify in the next subsection, there are well-founded reasons to expect this result. 

\section{Bloch Sphere Parameters for Meson-Antimeson Systems}

As an immediate generalisation of the results presented in Section~\ref{sec:Bmeson}, we provide a compilation table of all well-explored meson--antimeson two-level systems in terms of the Bloch-sphere parameters: $r$, $\theta_{\mathbf{e}\boldsymbol{\gamma}}$ and $|{\bf E}|$. 

\begin{table}[t!]
    \centering
    \begin{tabular}{|c||c|c|c|}
        \hline
        Meson & $\Delta E = \Delta m\ [\text{ps}^{-1}]$  & $\Delta
                                                           \Gamma\ [\text{ps}^{-1}]$
                                                        & $|q/p|-1$\\[0.1cm]
        \hline\hline
        $K^0$  & $0.005293\pm 9\times 10^{-6}$ & $0.01 \pm 5\!\times\!10^{-6}$ &
 $-0.003239\pm 10^{-6}$\\[0.1cm]
        \hline
        $D^0$&  $0.01\pm 0.001$ & $0.03 \pm  0.003$ & $(-5.00 \pm 0.04)\!\times\!10^{-3}$\\[0.1cm]
        \hline
        $B_{\rm d}^0$&  $0.5069 \pm 0.0019$ & $(0.7\pm 7)\!\times\!
                                              10^{-3}$ & $(1.0 \pm 0.8)\!\times\! 10^{-3}$\\
        \hline
        $B_{\rm s}^0$ &  $17.765\pm 0.006$ & $0.084\pm 0.005$ & $(0.1 \pm 1.4)\!\times\! 10^{-3}$\\[0.1cm]
        \hline
\end{tabular}
    \caption{\em Meson--Antimeson mixing data from the PDG~2024.}
    \label{tab:PDG2024}
\end{table}

\begin{table}[t!]
    \centering
    \begin{tabular}{|c||c|c|c|}
        \hline
        Meson & $r \equiv |{\bf \Gamma}|/(2|{\bf E}|)$ &                  $\theta_{\mathbf{e}\boldsymbol{\gamma}}[^\circ]$ &
        $|\mathbf{E}| [\text{ps}^{-1}]$\\[0.1cm]
        \hline\hline
                 $K^0$  & $0.945 \pm 2\times 10^{-3}$ & $179.6322\pm 1\!\times\!10^{-4}$&
        $2.64652 \times 10^{-3}\pm 7\!\times\!10^{-8} $
     \\[0.1cm]
        \hline
        $D^0$&  $1.5\pm 0.2$ & $179\pm 2$ & $(5.00 \pm 0.04)\!\times\!10^{-3}$\\[0.1cm]
        \hline
        $B_{\rm d}^0$&  $(1\pm 4) \times 10^{-3}$ & $-90 \pm 90$ & $0.253 \pm 0.001 $\\
        \hline
        $B_{\rm s}^0$ &  $(2.4 \pm 0.2) \times 10^{-3}$ & $182.7\pm 33.8$ & $8.9 \pm 0.1$\\[0.1cm]
        \hline
\end{tabular}
    \caption{\em Meson--Antimeson systems in terms of the Bloch-sphere parameters: $r$, $\theta_{\mathbf{e}\boldsymbol{\gamma}}$ and $|{\bf E}|$.}
    \label{tab:BlochMesons}
\end{table}

In Table~\ref{tab:PDG2024}, we present the current meson-antimeson mixing data from~\cite{ParticleDataGroup:2024cfk}. Here, we advocate that, in addition to these data, it would be further instructive to re-express them in terms of Bloch-sphere parameters, as done in Table~\ref{tab:BlochMesons}. It is interesting to note that, with a possible exception of the $B_{\rm d}^0$ system, all other known meson-antimeson systems have been found to have the Bloch vectors ${\bf E}$ and ${\bf \Gamma}$ well aligned. 

Finally, it is important to note that a meson-antimeson system 
with $r > 1$ describes an unstable qubit, which is over-damped and does not exhibit oscillation. Specifically, this is the case for the $D^0\bar{D}^0$ system, for which $r \simeq 1.5$ and as such, it does not oscillate, in complete agreement with experimental observations~\cite{ParticleDataGroup:2024cfk}.

\section{Conclusions}

The present study provides a complete analysis of CUQs, where we have detailed the analytic evolution of the unstable system, and whose time evolution is described by an infinite series of Fourier modes. However, there are still questions on quasi-critical unstable qubit scenarios where $\mathbf{e}$ is not quite orthogonal to $\boldsymbol{\gamma}$. Numerical evaluation of the master evolution equation seems to suggest that the oscillation period is solely dependent on $r$ and therefore independent of the angle between $\mathbf{e}$ and $\boldsymbol{\gamma}$, i.e. $\theta_{\mathbf{e}\boldsymbol{\gamma}}$. Future studies may consider how a similar Fourier-type decomposition can be performed to extract not only the dimensionless parameter $r$, but also~$\theta_{\mathbf{e}\boldsymbol{\gamma}}$. Consequently, meson--antimeson systems with quasi-degenerate decay widths can be included, thus providing an alternate method to detect CP violation and search for new physics. Since this study captures much of the physics pertinent to isolated quantum systems, a natural extension would be to consider these ideas in the context of entangled open quantum systems.

\section*{Acknowledgements}
The work of SP is funded by the Faculty of Science and Engineering Bicentenary Scholarship from Manchester U.,
and AP's work is supported in part by
the STFC research grant: ST/X00077X/1. 

\appendix

\section{Mixed States}\label{app:Mixed}

In the following section, we will map out a derivation for the evolution of the general mixed state CUQ. Initially, we will only consider the magnitude of the co-decaying Bloch vector, $\vecb(\tau)$, as a function of the angle between $\vecb(\tau)$ and the decay vector, $\boldsymbol{\gamma}$, i.e., $\varphi = \cos^{-1}\lrb{\vecb\cdot\boldsymbol{\gamma}/|\vecb|}$. However, taking this detour will provide us with some insights into the unique properties of the mixed state CUQ, as well as simplify some of the evolution equations from which we derive the analytic evolution of the mixed state CUQ as a function of $\tau$.

In \cite{Karamitros:2022oew} and \cite{Karamitros:2025azy}, it was shown explicitly that the motion of $\vecb$ in a CUQ is planar due to vanishing torsion. Moreover, a central feature of CUQ systems is indefinite oscillations whenever $r<1$. It is, therefore, appropriate to describe $\vecb$ using a polar representation, i.e., $|\vecb|$ and $\varphi = \cos^{-1}\lrb{\vecb\cdot\boldsymbol{\gamma}/|\vecb|}$. In the previous works~\cite{Karamitros:2022oew,Karamitros:2025azy}, it was demonstrated that when $\vecb(0)$ is confined to the plane spanned by $\lrcb{\boldsymbol{\gamma}, \mathbf{e}\times\boldsymbol{\gamma}}$, it will remain confined to this plane as the system evolves. Furthermore, it was proven that all CUQs are equivalent to this system up to an appropriate reparametrisation. Consequently, we are free to make use of the natural planar basis $\lrcb{\boldsymbol{\gamma}, \mathbf{e}\times\boldsymbol{\gamma}}$ in our analysis of mixed states. With these assumptions laid out, we will consider the projections:
\begin{equation}
    \vecb \cdot \boldsymbol{\gamma} \:  = \: |\vecb| \, \cos \varphi \, , \qquad \vecb \cdot \lrb{\mathbf{e}\times\boldsymbol{\gamma}} \:  = \: |\vecb| \, \sin \varphi \, .
\end{equation}

As previously introduced, the evolution of the co-decaying Bloch vector, $\vecb$, is fully described by the master evolution equation~\eqref{eq:bEvolEq2}. In order to study the evolution of $|\vecb|$ and $\varphi$, we make use of the following results:
\begin{subequations}\label{eq:MixStEvols}
    \begin{align}
    \frac{\de |\vecb|}{\de \tau} \: = \: \frac{1}{|\vecb|}\frac{\de \vecb}{\de \tau}\cdot \vecb \: = \: \lrsb{1-|\vecb|^2} \, \cos\varphi \, , 
\end{align}
\begin{align}
    \frac{\de}{\de \tau} \lrb{ \frac{\vecb \cdot \boldsymbol{\gamma}}{|\vecb|} } \: = \: \frac{1}{|\vecb|} \frac{\de \vecb}{\de \tau} \cdot \boldsymbol{\gamma} -  \frac{\vecb \cdot \boldsymbol{\gamma}}{|\vecb|^2}\frac{\de |\vecb|}{\de \tau} \: = \: \frac{1}{r}\sin\varphi + \frac{1}{|\vecb|}\sin^2\varphi \, .
\end{align}
\end{subequations}
Taken together, these expressions give the full planar evolution of both pure and mixed CUQ systems:
    \begin{equation}\label{eq:Polar_eqns}
        \frac{\de |\vecb|}{\de \tau}\: = \: \lrsb{1-|\vecb|^2} \, \cos\varphi \, , \qquad \frac{\de \varphi}{\de \tau} \: = \: -\frac{1}{r} - \frac{1}{|\vecb|}\sin\varphi \, .
    \end{equation}
Solving these equations directly to find the $\tau$-dependence of the general CUQ is, for the most part, not straightforward.

In the pure state special case where $|\vecb|=1$, we can see that the purity of states is preserved as the state evolves, and consequently, the full evolution of $\vecb$ may be reduced to a single differential equation for $\varphi$. This special case is rich in underlying physics, with current experiments on meson systems regularly making use of this assumption~\cite{LHCb:2016gsk, CDF:1996kan}. Additionally,~\cite{Karamitros:2025azy} considers these systems in great detail, highlighting novel features such as anharmonic oscillations, and identification of the oscillation frequency to be
\begin{equation}
    \widehat{\rm P}_{\rm pure} \: = \: \frac{2\pi r}{\sqrt{1-r^2}} \, .
\end{equation}

\begin{figure}
    \centering
    \includegraphics[width=0.5\linewidth]{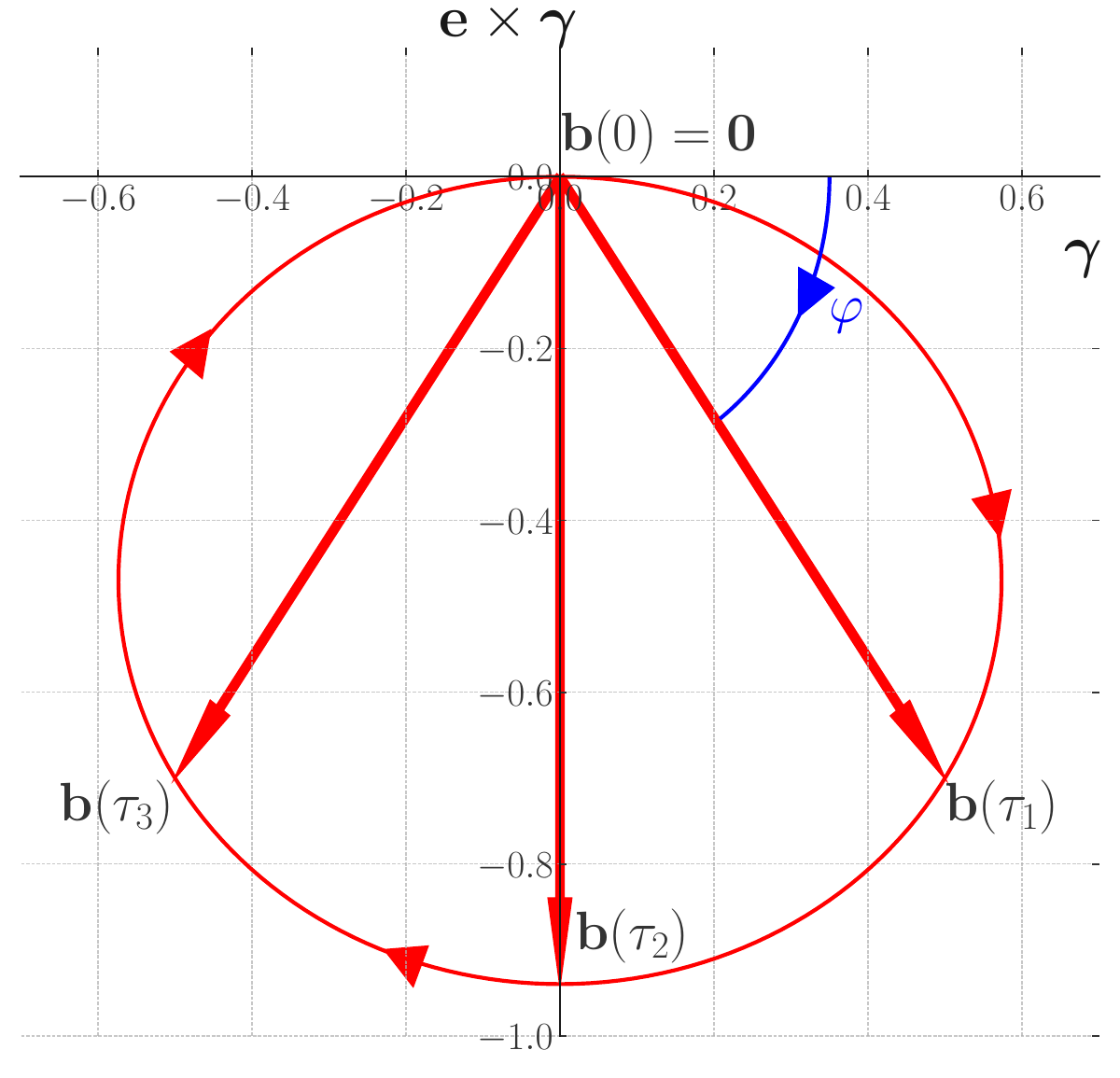}
    \caption{The evolution of $\vecb$ as a function of the angle between $\vecb$ and $\boldsymbol{\gamma}$, $\varphi = \theta + {\pi\over 2}$. The solid red line shows values of $\be(\theta)$ in the range $[-\pi, 0]$. Given the oscillation period $\widehat{\rm P}$, we define $\tau_1 = \frac{1}{4}\widehat{\rm P}$, $\tau_2 = \frac{1}{2}\widehat{\rm P}$, and $\tau_1 = \frac{3}{4}\widehat{\rm P}$.}
    \label{fig:magb_theta}
\end{figure}

Away from the pure state assumption, we see from this system of differential equations that the evolution is less simple, since both the magnitude of $\vecb$ and its angle with $\boldsymbol{\gamma}$ change simultaneously and interdependently. We do, however, notice that both evolution equations are fully autonomous, and consequently, may be used to find a compact expression that describes the magnitude of $\vecb$ as a function of $\varphi$. Taking the quotient of the expressions given in~\eqref{eq:MixStEvols} gives:
\begin{equation}
    \frac{\de |\vecb|}{\de \varphi} \: = \: \frac{\de |\vecb|}{\de\tau}\frac{\de \tau}{\de \varphi} \: = \: - \frac{r |\vecb| \lrsb{1-|\vecb|^2}\cos\varphi}{|\vecb|+r\sin\varphi} \, .
\end{equation}
The solution for this new expression remains awkward to attain, however, may be done by writing it as an exact differential, and dividing through with the integrating factor $\sqrt{1-|\vecb|^2}$. This gives us an exact differential of the form:
\begin{equation}
    \dfrac{|\mathbf{b}|+r\sin{\varphi}}{(1-|\mathbf{b}|^2)^{3/2}}\mathrm{d}|\mathbf{b}| + \dfrac{r|\mathbf{b}|\cos{\varphi}}{\sqrt{1-|\mathbf{b}|^2}} \mathrm{d}\varphi = 0.
\end{equation}
For our purposes, we are predominantly interested in the case where the initial condition of the CUQ is a fully mixed state, i.e., $|\vecb(0)|=0$. Applying this condition, one sees that the magnitude of $\vecb$ takes the following analytic form:
\begin{equation}\label{eq:Magbfunc}
    |\vecb| \: = \: - \, \frac{2r\sin\varphi}{1+r^2 \sin^2\varphi} \: = \: \frac{2r\sin(\varphi+\pi)}{1+r^2 \sin^2(\varphi+\pi)} \, .
\end{equation}
Whilst simple, this expression contains an unexpected result. Careful consideration of the right-hand side shows that the periodicity of $|\vecb|$ is not truly $\varphi \in [0,2\pi]$, but compactified into the smaller range $\varphi \in [0, \pi]$. As a result, over a single rotation period for $\varphi$, for a fully mixed initial state, $\vecb(\tau)$ will, in fact, only sweep out a path in the lower half plane. This is shown in Figure~\ref{fig:magb_theta}.

In closing, we see that there is a neat form for the path traced out by the co-decaying Bloch vector, $\be$. Using the expression for $|\be|$, we find that the angular evolution of the projections is given by:
\begin{subequations}
    \begin{align}
    \vecb \cdot \boldsymbol{\gamma} \: = \: |\vecb|\cos\varphi \: = \: - \frac{2r\sin\varphi \cos\varphi}{1+r^2\sin^2\varphi} \, , \\
    \vecb \cdot \lrb{\mathbf{e}\times\boldsymbol{\gamma}} \: = \: |\vecb|\sin\varphi \: = \: - \frac{2r\sin^2\varphi}{1+r^2\sin^2\varphi} \, .
\end{align}
\end{subequations}
We know that ${\rm max}|\be| = 2r/(1+r^2)$, and that this occurs at $\varphi=-\pi/2$. Furthermore, the symmetry properties of $|\be(\varphi)|$ indicate that the path traced out by $\be$ is symmetric along $\boldsymbol{\gamma}$. Consequently, we posit that the centre of the path lies along the $\lrb{\mathbf{e}\times\boldsymbol{\gamma}}$ direction. We assume that the path along this direction has its own symmetry properties derived from the symmetries of the trigonometric functions, and so we centre the path on the origin by shifting the projection along $\mathbf{e}\times\boldsymbol{\gamma}$ by half the maximum. We hence introduce the shorthand notations:
\begin{subequations}
    \begin{align}
        {\rm X} \: &= \: - \frac{2r\sin\varphi \cos\varphi}{1+r^2\sin^2\varphi} \, , \\
        {\rm Y} \: &= \: - \frac{2r\sin^2\varphi}{1+r^2\sin^2\varphi}-\frac{r}{1+r^2}.
    \end{align}
\end{subequations}
By utilizing some trigonometric identities, we can obtain the trajectory of the co-decaying Bloch vector as
\begin{equation}
    \lrsb{\frac{\sqrt{1+r^2}}{r} {\rm X}}^2 + \lrsb{\frac{1+r^2}{r} {\rm Y}}^2 \: = \: 1 \, .
\end{equation}
Hence, the path of $\be$ is similar to an ellipse with semi-major and semi-minor axes:
\begin{equation}
    {\rm Y}^* \: = \: \frac{r}{1+r^2} \, .
\end{equation}
\begin{equation}
    {\rm X}^* \: = \:  \frac{r}{\sqrt{1+r^2}} \, .
\end{equation}
As a result, we assume that the semi-major axis of the ellipse is given by ${\rm X}^*$. Finally, we restore the original definitions of ${\rm X}$ and ${\rm Y}$ to write these expressions in terms of the original projections of $\be$:
\begin{equation}
    \lrsb{\frac{\sqrt{1+r^2}}{r} \vecb \cdot \boldsymbol{\gamma}}^2 + \lrsb{1 + \frac{1+r^2}{r} \vecb \cdot \lrb{\mathbf{e}\times\boldsymbol{\gamma}}}^2 \: = \: 1 \, .
\end{equation}
Given the semi-minor axis, ${\rm Y}^*$, and semi-major axis, ${\rm X}^*$, one can then measure the eccentricity of the mixed CUQ to find:
\begin{equation}
    e^2 \: \equiv \: 1 - \lrb{\frac{{\rm Y}^*}{{\rm X}^*}}^2 \: =\: 1 - \lrb{\frac{1}{1+r^2}} \: = \: \frac{r^2}{1+r^2} \, .
\end{equation}
For a small value of $r$, we can approximate the measure of eccentricity:
\begin{equation}
    e^2 \: \simeq \: r^2 + \mathcal{O}(r^4).
\end{equation}
We then see the remarkable result that the eccentricity is itself a measure of $r$, i.e., $e \simeq r$. Indeed, we see that as $r$ approaches zero, both the semi-minor and semi-major axes approach $r$. Consequently, the ellipse approaches a circle of radius $r$, and the eccentricity drops to zero.

\begin{figure}[t!]
    \centering
    \includegraphics[scale=0.3]{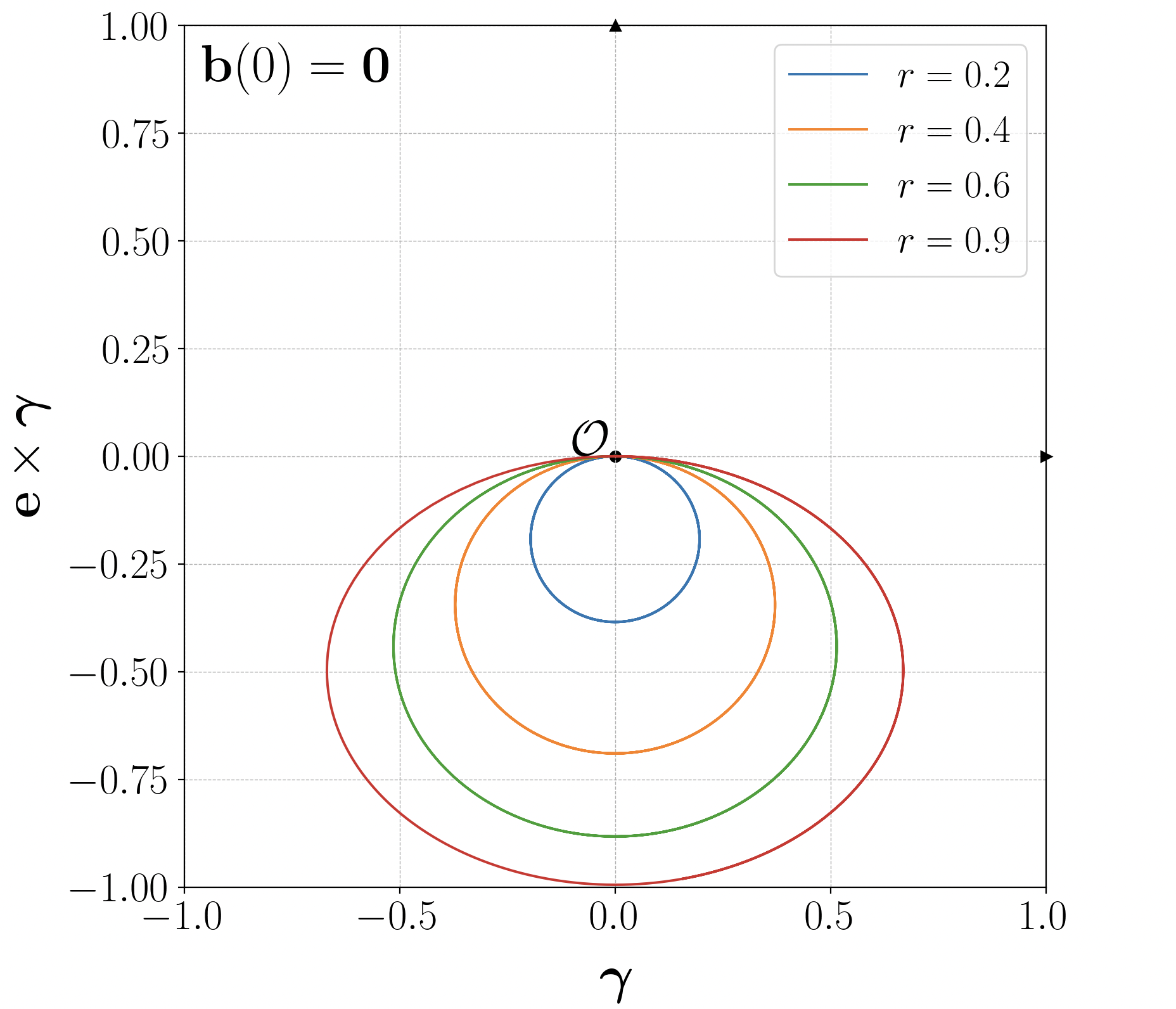}
    \caption{Elliptical trajectories for the mixed state CUQ. The exact solution of the master evolution equation given the initial condition $\be(0)=\mathbf{0}$ are plotted for $r=0.2$, $r=0.4$, $r=0.6$ and $r=0.9$.}
    \label{fig:Ellipses}
\end{figure}

\vfill\eject

\bibliography{bibs-refs}{}
\bibliographystyle{unsrt}

\end{document}